\documentstyle[preprint,eqsecnum,aps]{revtex}
\tightenlines

\begin{document}
\draft
\preprint{HEP/123-qed}
\title{
Relation between the superconducting gap energy and the two-magnon 
Raman peak energy in Bi$_2$Sr$_2$Ca$_{1-x}$Y$_x$Cu$_2$O$_{8+\delta}$
}

\author{S. Sugai and T. Hosokawa}
\address{Department of Physics, Faculty of Science, Nagoya University, 
Chikusa-ku, Nagoya 464-8602, Japan} 

\date{\today}
\maketitle

\begin{abstract}
The relation between the electronic excitation and the magnetic 
excitation for the superconductivity in 
Bi$_2$Sr$_2$Ca$_{1-x}$Y$_x$Cu$_2$O$_{8+\delta}$ was investigated 
by wide-energy Raman spectroscopy.  
In the underdoping region the $B_{\rm 1g}$ scattering intensity is 
depleted below the two-magnon peak energy due to the "hot spots" 
effects.  
The depleted region decreases according to the decrease of the 
two-magnon peak energy, as the carrier concentration increases.  
This two-magnon peak energy also determines the $B_{\rm 1g}$ 
superconducting gap energy as $2\Delta \approx \alpha \hbar 
\omega_{\rm Two-Magnon} \approx J_{\rm effective}$ 
$(\alpha=0.34-0.41)$ from under to overdoping hole concentration.  
\end{abstract}


\narrowtext
	The effects of strong electron-spin interactions in high 
$T_{\rm C}$ superconductors manifest themselves both in charge excitation 
spectra and spin excitation spectra.  
For example, the angle-resolved photoemission spectroscopy (ARPES) 
revealed that electronic states near $(\pi, 0)$ ("hot spots") are 
depleted in the underdoping region, because large parts of the 
electronic states become incoherent as the result of interactions with
electrons near $(0, \pi)$ via collective magnetic excitations 
near $(\pi, \pi)$ \cite{1,2,3}.  
The resonance peak energy of this magnetic excitation, which is observed 
in the neutron scattering spectroscopy, appears in ARPES as the energy 
difference between the peak and the dip in the superconducting 
gap structure \cite{4,5,6,7}.  
Raman scattering can measure both charge and spin excitations 
simultaneously in the superconducting state, because the two-magnon 
scattering persists even in the metallic states \cite{8,9,10,11}.  
The present experiment aims to obtain the relation among the two-magnon 
scattering energy, which is the measure of the effective exchange energy, 
the superconducting gap energy, and the boundary energy of the 
depleted $B_{1g}$ scattering intensity in the underdoping region.

	Electronic Raman scattering can detect selected parts in the 
$k$-space by choosing the combination of the incident and scattered 
light polarizations \cite{12,13,14,15}.  
When the polarizations of the incident and scattered light are parallel 
to the $x$ and $y$ directions, respectively, ($(xy)$-polarization 
configuration), the allowed symmetry is the $B_{\rm 1g}$ and the 
observable electronic excitations are mainly near the region 
connecting (0, 0) and $(\pi,\ 0)$ in the $k$-space.  
The $d(a^2-b^2)$ superconducting gap has the maximum in the 
(0, 0)-$(\pi,\ 0)$ direction.  
Here $x$ and $y$ axes are at $45^{\circ}$ from the $a$ and $b$ axes 
which are along the direction connecting Cu-O-Cu.   
In the $(ab)$-polarization configuration, the allowed symmetry is 
$B_{\rm 2g}$ and the electronic excitations mainly near 
(0, 0)-$(\pi,\ \pi)$ are detected.  
The $d(a^2-b^2)$ superconducting gap has the node in this direction.  
The observed gap energy is larger in $B_{\rm 1g}$ than in $B_{2g}$ 
consistently with the $d(a^2-b^2)$ superconductivity \cite{12,16}.  
The $B_{\rm 1g}$ gap energy increases, as the carrier concentration 
decreases \cite{17,18,19,20,21}.  
While the $B_{\rm 2g}$ gap energy follows the $T_{\rm C}$ \cite{22,23}.  
As going to the overdoping region, the $B_{\rm 2g}$ superconducting 
gap approaches to the $B_{\rm 1g}$ gap in energy looking like a loss of 
the anisotropy \cite{20}, while the ARPES observed the clear gap 
node \cite{24}.  
The $B_{\rm 1g}$ gap structure as well as the $B_{\rm 1g}$ scattering 
intensity itself decreases, as the carrier concentration decreases 
\cite{25,26,27,28}.  
It can be attributed to the "hot spots" effects \cite{1,2,3}.  
Many Raman scattering studies on 
Bi$_2$Sr$_2$Ca$_{1-x}$Y$_x$Cu$_2$O$_{8+\delta}$ have been performed 
\cite{8,10,11,12,17,19,20,21,22,23,yamanaka,quilty}. 
However, the relation among the effective exchange energy obtained 
from the two-magnon scattering, the energy width of the incoherent 
electronic states estimated from the depleted electronic scattering 
intensity, and the superconducting gap energy has not been reported.

	Single crystals of Bi$_2$Sr$_2$Ca$_{1-x}$Y$_x$Cu$_2$O$_{8+\delta}$ 
were synthesized by the travelling solvent floating zone method 
utilizing an infrared radiation furnace with quaternary oval mirrors.  
The starting composition for the feed and seed rods is 
Bi$_{2.1}$Sr$_{1.9}$Ca$_{1-x}$Y$_x$Cu$_2$O$_{8+\delta}$ and that for 
the solvent is Bi$_{2.2}$Sr$_{1.6}$Ca$_{0.85}$Cu$_{2.2}$O$_{z}$ 
\cite{mochiku}.
The crystals used in this experiment are an overdoped sample 
($x=0$, mid-point of the transition $T_{\rm C}$=84 K, hole 
concentration $/$Cu $p=0.20$), an optimally doped sample 
($x=0.1$, $T_{\rm C}$=95 K, $p=0.16$), and underdoped samples 
($x=0.2$, $T_{\rm C}$=87 K, $p=0.13$ and $x=0.3$, 
$T_{\rm C}$=75 K, $p=0.11$).  
Here $p$ is estimated using 
$T_{\rm C}/T_{\rm C}^{\rm max}=1-82.6(p-0.16)^2$ with 
$T_{\rm C}^{\rm max}=95$ K \cite{29}.

	Raman spectra were measured on fresh cleaved surfaces in a 
quasi-back scattering configuration utilizing a triple monochromator, 
a liquid nitrogen cooled CCD detector, a 5145 \AA \ Ar-ion laser.  
The laser beam was focused on the area of 50 $\mu$m$\times$500 $\mu$m.  
The wide-energy spectra were measured at the laser power 25 mW and 
the low-energy spectra at 10 mW.  
The increase of temperature was less than 2 K for the 10 mW excitation.  
The same spectra were measured four times to remove the cosmic ray 
noise by comparing the intensities at each channel.  
The wide energy spectra covering $12-7000$ cm$^{-1}$ was obtained by 
connecting 45 spectra with narrow energy ranges after correcting 
the spectroscopic efficiency of the optical system.  
The same spot on the surface was measured during the temperature 
variation by correcting the sample position viewed through a TV camera 
inside the spectrometer.  
The experimental fluctuation for the intensity is less than $\pm 5$ 
\% throughout the present experiments.

	Figure 1 shows the $B_{\rm 1g}$, $B_{\rm 2g}$ and $A_{\rm 1g}$ 
Raman spectra in the superconducting state at 20 k and in the normal 
state at 100 K.  
The $A_{\rm 1g}$ spectrum is obtained by subtracting the $(xy)$ 
spectrum from the $(aa)$ spectrum.  
As the hole concentration decreases, the $B_{\rm 1g}$ two-magnon 
peak shifts to higher energy and connects with the two-magnon energy 
3050 cm$^{-1}$ in the antiferromagnetic insulator 
Bi$_2$Sr$_2$Ca$_{0.5}$Y$_{0.5}$Cu$_2$O$_{8+\delta}$ \cite{sugai}.  
The intensity of the $B_{\rm 1g}$ electronic scattering from 150 to 650 
cm$^{-1}$ at $x=0$ is  larger than the intensity of the two-magnon peak 
at 1000 cm$^{-1}$.  
The low-energy electronic scattering intensity below the two-magnon 
peak energy decreases, as the hole concentration decreases.  
This depletion of the scattering intensity is consistent with the 
depletion of the low-energy spectral intensity at the "hot spots" 
near $(\pi,\ 0)$ observed by ARPES \cite{1,2,3}.  
The finding that the $B_{\rm 1g}$ Raman intensity is depleted 
below the two-magnon peak energy indicates that the electron 
self-energy includes the process of transition from $(\pi,\ 0)$ to 
$(0,\ -\pi)$ by emitting the $(\pi,\ \pi)$ magnon and back to 
$(\pi,\ 0)$ by emitting the $(-\pi,\ -\pi)$ magnon.  
The creation of the $(\pi,\ \pi)$ and $(-\pi,\ -\pi)$ magnons is the 
same as the process of the two-magnon Raman scattering peak.  
In the superconducting state at 20 K, the gap structure is strongly 
enhanced, as the hole concentration increases to the optimum and the 
overdoping region.

	The hole concentration dependence of the $B_{\rm 2g}$ spectrum 
is quite different from the $B_{\rm 1g}$ spectrum.  
In the 100 K spectrum at x=0.3, the scattering intensity decreases 
monotonically from 200 cm$^{-1}$ to over 7000 cm$^{-1}$.  
As hole concentration increases above $x=0.2$, the scattering intensity 
decreases below 1500 cm$^{-1}$ in the form of a step function.  
The drop of the intensity is observed near 4000 cm$^{-1}$ besides the 
step-like decrease at 20 K.  
The interaction of electrons at $(-\pi / 2,\ -\pi /2)$ and at 
$(\pi / 2,\ \pi /2)$ via the $(\pi,\ \pi)$ magnetic excitation may 
contribute these structure.  
It may be noted that the 1500 cm$^{-1}$ is the upper limit of the strong 
resonant two-phonon scattering.  
At 20 K, the gap structure appears with almost the same strength for 
all samples.

	The $A_{\rm 1g}$ spectrum shows the intermediate carrier 
concentration dependence between the $B_{\rm 1g}$ and $B_{\rm 2g}$ spectra.  
The electronic Raman intensity decreases below 1300 cm$^{-1}$ at $x=0.3$.  
The low-energy scattering intensity increases, as the carrier 
concentration increases.

	Figure 2 shows the $B_{\rm 1g}$, $B_{\rm 2g}$, and $A_{\rm 1g}$ 
[$(xx)$-$(ab)$] spectra at 20 K and 100 K.  
It is obvious that the gap structure is strongly enhanced at the 
optimum and the overdoped samples in the $B_{\rm 1g}$ and $A_{\rm 1g}$ 
spectra, but the intensity is almost the same in the $B_{\rm 2g}$ 
spectra.  
In order to eliminate the phonon peaks, the difference spectra between 
at 20 K and at 100 K are plotted in Fig. 
3.  The peak energy is assigned to the superconducting gap energy.  
The gap energy increases in $B_{\rm 1g}$ and $A_{\rm 1g}$, as the hole 
concentration decreases, but decreases in $B_{\rm 2g}$ except for 
the small increase from $x=0$ to $x=0.1$.

	The energies of the superconducting gap and the two-magnon peak 
at 20 K are plotted in the upper panel of Fig. 4.  
Integrated relative intensities of the superconducting gap peak for 
$I(20 {\rm K})>I(100 {\rm K})$ are plotted in the lower panel of Fig. 4.

	The $k$-dependent gap structure was investigated by 
ARPES \cite{28,30}.  
The results are (1) the gap energy on the Fermi surface has a node 
on the (0, 0)-$(\pi,\ \pi)$ line irrespective of the hole concentration, 
(2) the maximum gap energy on the (0, 0)-$(\pi,\ 0)$ direction increases, 
as hole concentration decreases, (3) the angular-dependent gap energy 
from the (0, 0)-$(\pi,\ \pi)$ direction to the (0, 0)-$(\pi,\ 0)$ direction 
changes from the linearly increasing function of the angle near 
(0, 0)-$(\pi,\ \pi)$ in the optimum and overdoped samples to the function 
with positive curvature in the underdoped samples.  
The hole concentration dependence of the $B_{\rm 1g}$ and $B_{\rm 2g}$ 
gap energies observed by Raman scattering is consistent with the 
results of ARPES, that is, the energy of the $B_{\rm 1g}$ gap which 
represents mainly the gap near (0, 0)-$(\pi,\ 0)$ increases, as the hole 
concentration decreases, and the energy of the $B_{\rm 2g}$ gap 
which represents the gap near (0, 0)-$(\pi,\ \pi)$ decreases.  
However, the following experimental results cannot be explained in 
the simple picture.  The first is that the $B_{\rm 2g}$ gap energy 
is almost the same as the $B_{\rm 1g}$ gap energy in the overdoped 
sample at x=0, although the ARPES clearly observed the node on the 
(0, 0)-$(\pi,\ \pi)$ direction \cite{28}.  
The second is the $A_{\rm 1g}$ gap energy which is predicted to be 
inbetween the $B_{\rm 1g}$ and $B_{\rm 2g}$ gap energies differently 
from the experimental results.  
The theory noted that the gap energy observed by Raman scattering is 
sensitive to the structure of the Fermi surface \cite{12,13,14,15}.  
Further theoretical investigation is expected.

	The important point shown in Fig. 4 is that the $B_{\rm 1g}$ gap 
energy is proportional to the $B_{\rm 1g}$ two-magnon energy 
\begin{eqnarray}
2\Delta (B_{1 \rm g})=\alpha \hbar \omega(B_{1 \rm g} \ 
{\mbox{two-magnon}}), 
\end{eqnarray}
for the wide carrier concentration region from underdoping to 
overdoping.  
The proportionality coefficient $\alpha$ is about 0.4, gradually 
increasing from 0.34 at $x=0$ to 0.41 at $x=0.3$.
The two-magnon peak energy is about 3$J$ in the insulating phase, 
where $J$ is the exchange energy between Cu atoms.  
If this relation holds into the metallic phase, the energy gap 
equals about the effective exchange energy.  
In addition the upper limit energy for the depletion in the 
$B_{\rm 1g}$ spectrum is just the two-magnon peak energy in the 
underdoping region.  
These experimental results indicate that the superconductivity is 
directly induced by the interaction with the magnetic excitation at 
$(\pi,\ \pi)$.  
As for the depletion picture in the underdoping region near the 
insulator-metal transition, we can present the example where there 
is no depletion in both $B_{\rm 1g}$ and $B_{\rm 2g}$ spectra.  
BaCo$_{1-x}$Ni$_x$S$_2$ is the case, where the spectrum like the 
$B_{\rm 2g}$ 
spectrum of Fig. 1 appear abruptly both in $B_{\rm 1g}$ and 
$B_{\rm 2g}$ spectra, when the phase changes into the paramagnetic 
metallic state $(x>0.22)$ from the antiferromagnetic state 
$(x<0.22)$ \cite{31}.  
This abrupt increase of the electronic scattering intensity at the 
transition to the paramagnetic metallic phase is related to the 
enhancement of the electronic specific heat ($T$-linear coefficient 
$(\gamma)$ of the low-temperature specific heat) in the metallic 
phase near the transition.  
It is known that the antiferromagnetic insulator-metal transition 
in high $T_{\rm C}$ superconductors is characterized by no 
enhancement of $\gamma$ \cite{32}.  
Thus it can be concluded that $\gamma$ is not enhanced at the 
insulator-metal transition, if the electronic Raman scattering 
intensity, or the electronic density of states, is depleted by the 
"hot spots" effects, and vice versa.

	In conclusion the present experiment elucidates the relation 
among the $B_{\rm 1g}$ two-magnon peak energy, the $B_{\rm 1g}$ 
superconducting gap energy, and the upper-limit energy of the 
depleted electronic density of states near $(\pi,\ 0)$ due to 
the "hot spots" effects.  
These experimental results indicate that the $(\pi,\ \pi)$ magnetic 
excitation plays the crucial role for the high $T_{\rm C}$ 
superconductivity.

Acknowledgments - 
The authors would like to thank K. Takenaka for the 
characterization of single crystals.  
This work was supported by CREST of the Japan Science and 
Technology Corporation.

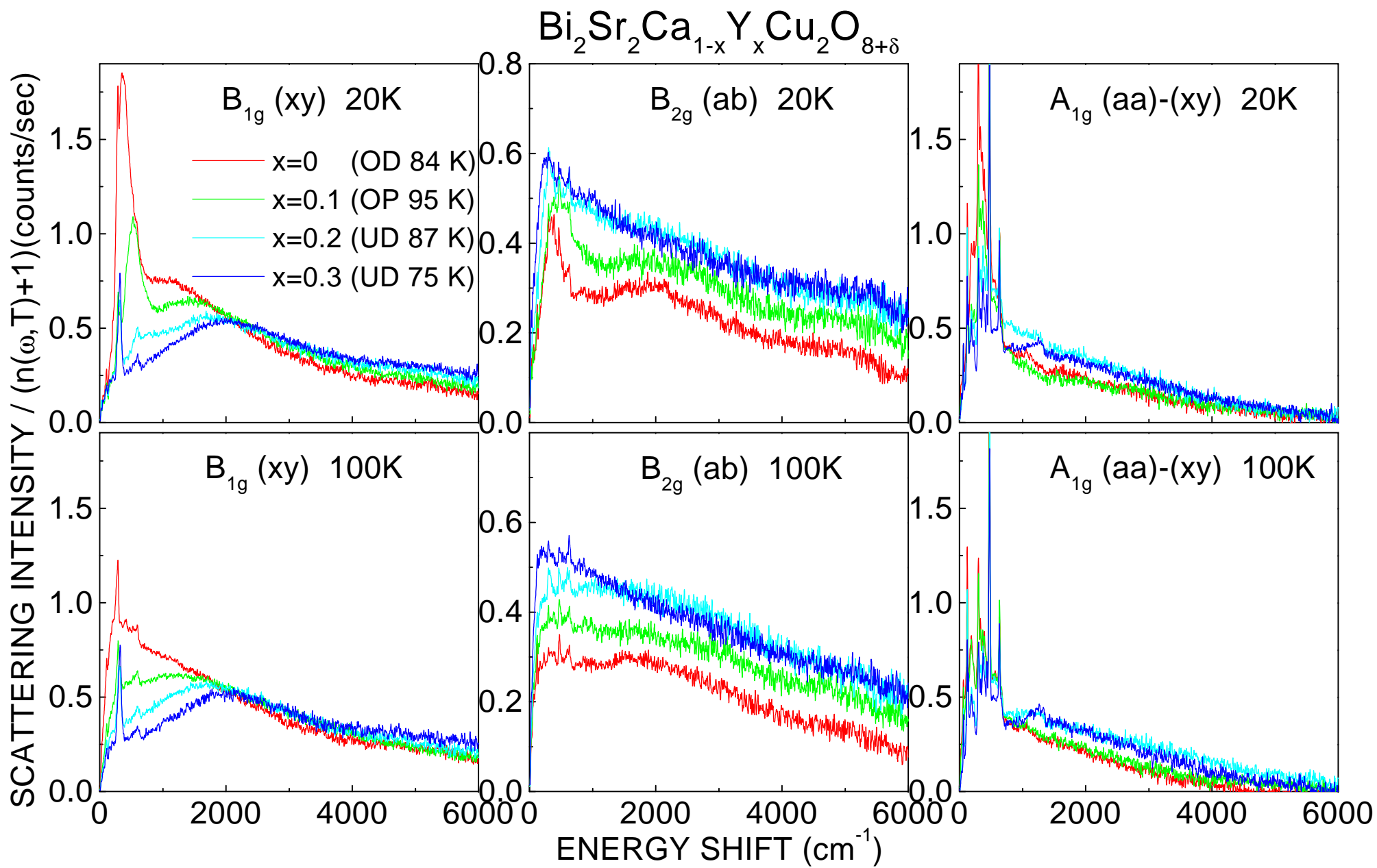
\begin{figure}
\caption[]{(color)
Carrier concentration dependence of the $B_{\rm 1g}$, $B_{\rm 2g}$, 
and $A_{\rm 1g}$ $[(aa)-(xy)]$ Raman spectra in the superconducting state 
at 20 K and in the normal state at 100 K.
}
\label{fig1}
\end{figure}

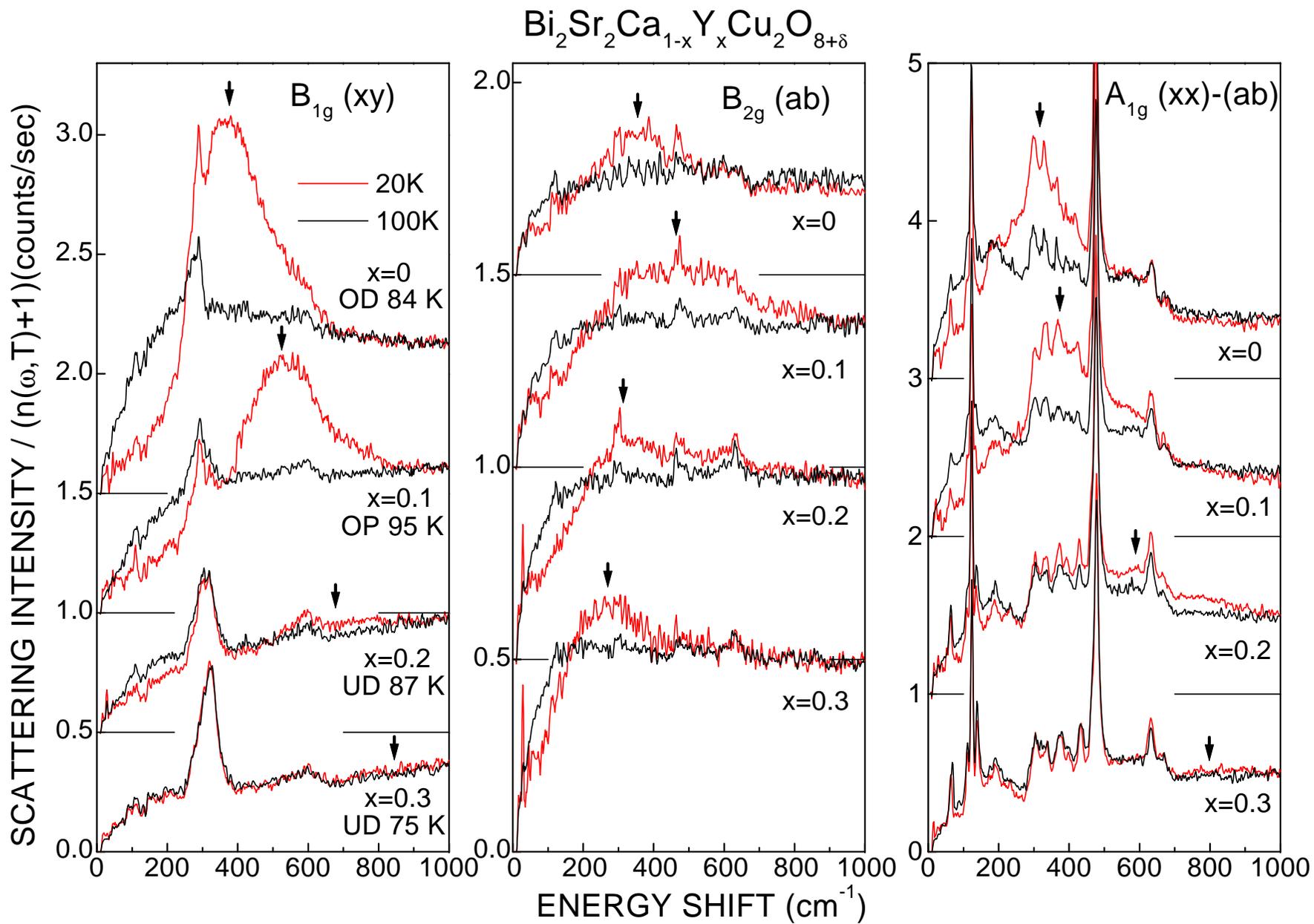
\begin{figure}
\caption[]{(color)
Superconducting gap structure in the $B_{\rm 1g}$, $B_{\rm 2g}$, 
and $A_{\rm 1g}$ $[(xx)-(ab)]$ Raman spectra.  
The difference peak positions shown in Fig. 3 are indicated by arrows.
}
\label{fig2}
\end{figure}

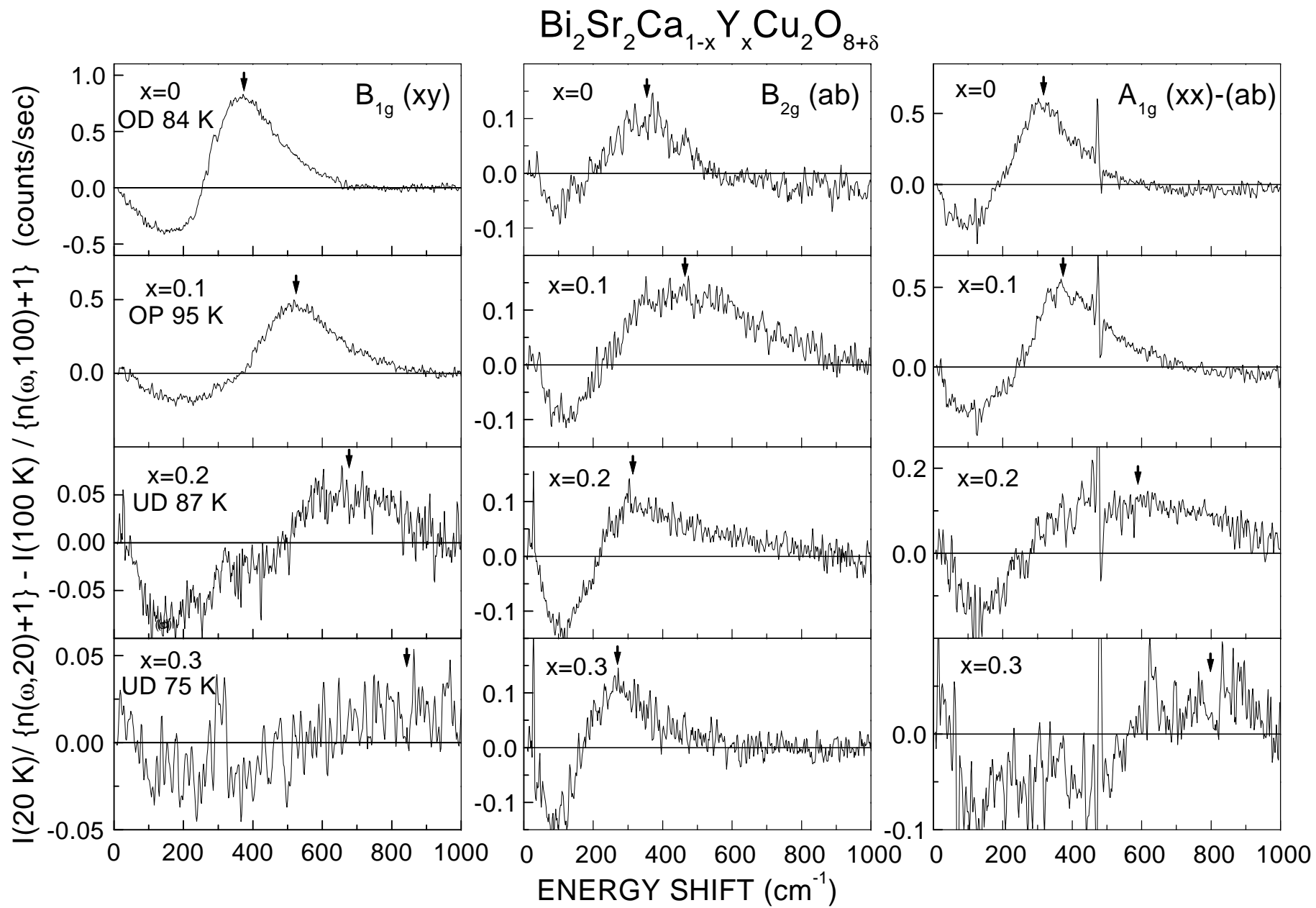
\begin{figure}
\caption[]{
The difference spectra between at 20 K and at 100 K.  
The peak positions are shown by arrows.
}
\label{f3}
\end{figure}

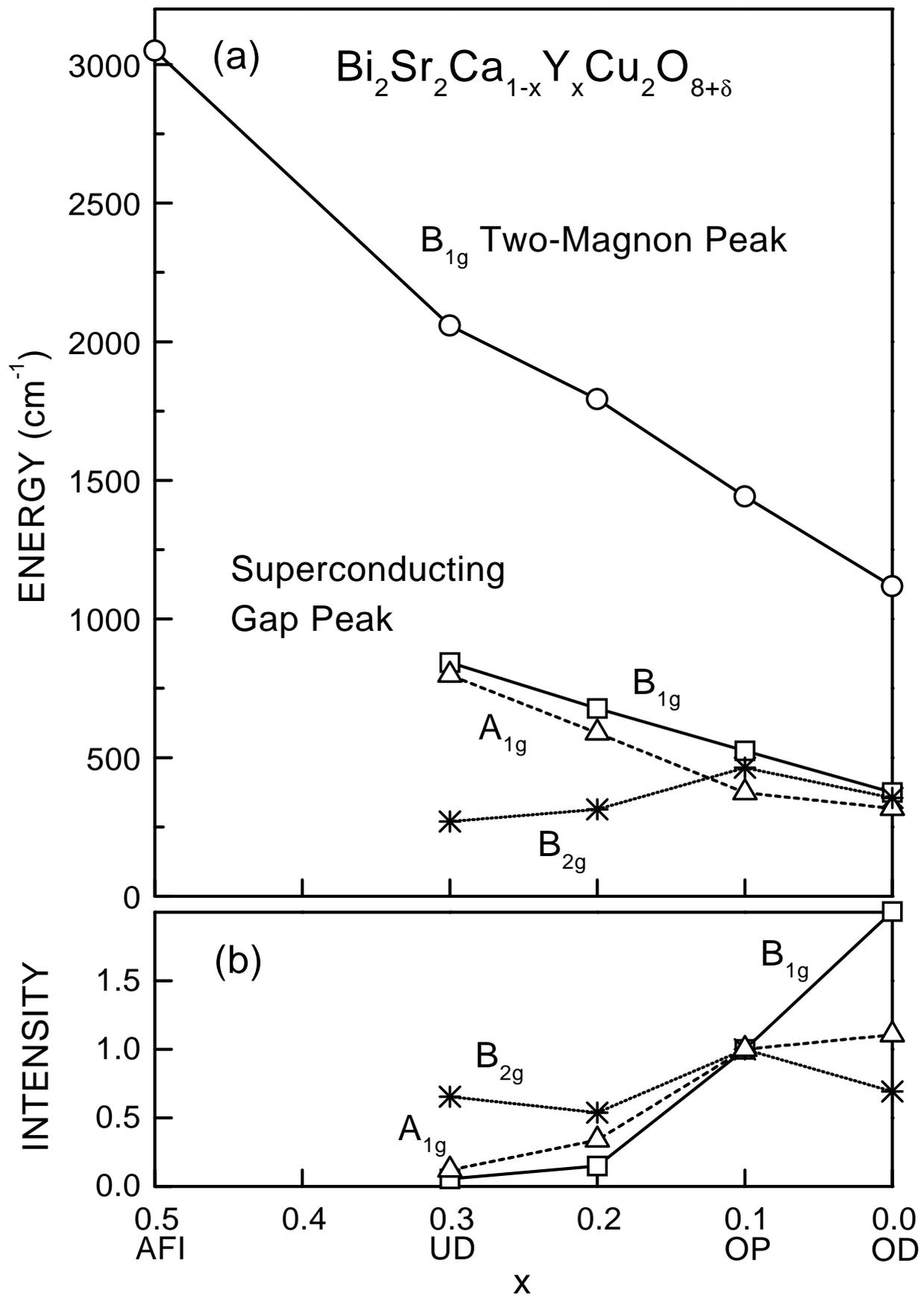
\begin{figure}
\caption[]{
(a) The two-magnon peak energy, the superconducting gap energy and 
(b) the relative integrated intensity of the superconducting gap peak 
for $I({\rm 20 K})>I({\rm 100 K})$ as a function of the carrier 
concentration.
}
\label{fig4}
\end{figure}

\end{document}